# A Comprehensive Study of Charge-Carrier Mobility in Double Lateral-Gate Silicon Junctionless Transistors

Farhad Larki[1,2,3], Arash Dehzangi[4]

[1]Laye Rooyan Part, Isfahan Science and Technology Town, Isfahan, 84156, Iran.
[2]Department of Physics, Isfahan University of Technology, Isfahan 84156, Iran.
[3]Institute of Microengineering and Nanoelectronics, Universiti Kebangsaan Malaysia, Bangi, Selangor, Malaysia.
[4]Department of Electrical and Computer Engineering, University of Texas at Dallas, Richardson, Texas, USA.
*arash.dehzangi@utdallas.edu

*Abstract*— **We give in depth overview of charge-carrier mobility in p-type silicon double lateral-gate junctionless transistors (DLGJLT) through a mix of computer simulations and experimental results. The devices were built on low-doped silicon-on-insulator substrates using atomic force microscope nanolithography. Because the design uses air-gap side gates instead of a gate-oxide layer or junction, it allows electrical current to flow through a lightly doped silicon channel. This setup reduces impurity scattering and protects transport quality from interface damage. We used three-dimensional TCAD Sentaurus simulation tool to model the device's electrical behavior and confirm our experimental data for transconductance and drain conductance. We attempt to give a comprehensive view on the charge transition within the device.**

***Keywords—Junctionless transistor, double lateral gate, hole mobility, hole velocity, effective mass***

## I. Introduction

Junctionless transistors (JLTs) achieved recent interests since they address several scaling and fabrication challenges faced by conventional Si based metal–oxide–semiconductor field-effect transistor (MOSFETs) in the nanometer regime. In traditional transistors, forming ultra-shallow, well-controlled source and drain junctions become increasingly difficult and expensive [1]. Junctionless devices eliminate these junctions entirely by using a uniformly doped semiconductor channel, meaning no distinctive region to be marked as source or drain junction in traditional manner [2-5].

By eliminating junction formation, junctionless transistors offer several advantages for future nanoelectronics, including lower fabrication complexity, reduced variability, and potentially improved scalability for future technology nodes [4, 6, 7].

They also offer strong electrostatic control and reduced short-channel effects, especially when implemented with nanowire or gate-all-around structures [6, 8-10]. Their simpler architecture, reduced variability, and compatibility with advanced 3D integration can make them an attractive candidate for next-generation low-power, high-density electronics. In the last decade, different approaches were suggested to enhance the operational efficiency of JLT devices. Various devices design such as FinFET, gate-all-around (GAA), single gate JLT, double gate JLT, thin film transistor (TFT), and tunnel FET (TFET) with various channel materials at different operating temperature have been investigated [11-25].

Due to the potential of junctionless transistors to improve carrier mobility through reduced junction-related scattering, investigating charge transport and mobility mechanisms in these devices is important for the development of reliable nanoscale junction-free transistors. This work presents a comprehensive view of charge-carrier mobility in p-type double lateral-gate junctionless transistor (DLGJLT) fabricated by scanning probe-based lithography (SPL). 3D TCAD Sentaurus simulations are used to analyze the electrostatic behavior of the structure. Variations in key physical dimensions, e.g. channel width, channel thickness, effective gate gap, and source/drain extension length modify the spatial localization of majority carriers, influence drift velocity, and reshape mobility profiles across the channel. We focus on the hole velocity and mobility of the majority carriers in a p-type DLGJLT device. This device has a unique structure with lateral gates separated by airgap from the channel. The fabrication process was reported somewhere else along with the device performance and charge transmission [26-28]. The device works based on pinch-off mechanism, which allows it to work with low doping profile. We also address the field effect, effective mobility and mobility improvement in accordance with low doping profile and particular shape of the device.

TCAD Simulation results were analyzed to evaluate the impact of channel width, gate gap, channel thickness and source/drain extension variation on carriers' mobility.

## II. Device Fabrication and Simulation

SPL is an alternative to conventional nanofabrication techniques, suggested mostly in academia for nanoscale device fabrication with versatility, flexibility, low cost, and nanoscale resolution. We used atomic force microscope (AFM) nanolithography to perform SPL on SOI to create facile and simple double lateral gate junctionless transistor (DLGJLT) devices [29, 30]. In this method, local anodic oxidation (LAO) process [31, 32] was implemented to define the mask on SOI, which is useful to maintain crystalline integrity of Si channel compared to other techniques, such as electron beam lithography (EBL). In this approach AFM tips are implemented to draw the oxidation pattern on prepared SOI substrate and has been extensively studied by several teams with acceptable repeatability [33-35].

LAO-AFM nanolithography in contact mode was used to fabricate the p-type double lateral gates Si nanowire JLT using scanning probe microscope (SPM) machine (SPI3800N/4000). The nanometer-scaled mask design was patterned on (100) SOI (SOI-SOITEC, 14-22 Ω cm, Boron doped) wafer. The upper Si thickness is 100 nm and buried oxide layer (BOX) is 200 nm. In LAO process a conductive AFM tip (Cr/Pt coating tip) was used to draw the oxidation pattern on prepared SOI substrate. During the LAO process and wet etching, all the involving parameters were studied and optimized to deliver a controllable process [32, 36].

To acquire accurate simulation of the device, we used a hydrodynamic model in Synopsys Sentaurus device simulator [37] as the platform for the 3-D TCAD simulation presented in this study.

The hydrodynamic transport model was adopted instead of the conventional drift–diffusion approach in order to capture non-equilibrium high-field transport effects, including carrier heating and velocity saturation, which become significant in the pinch-off regime of the DLGJLT structure. It should also be noted that no explicit quantum confinement or surface-roughness-scattering mobility model was included in the TCAD simulations. Therefore, discussions related to interface scattering and confinement effects are presented as qualitative physical interpretations rather than direct simulation-extracted quantities. The details for simulation parameters are given elsewhere [38, 39]. Along with the fundamental equations, the default Shockley–Read–Hall (SRH) recombination-generation model and doping-dependent Masetti mobility model were also considered. In all simulation steps, consistent with our experimental set up with a tungsten AFM tip, the contact work function is set to 5.12 eV. Given the device's dimensions range investigated, quantum confinement effects are not dominant under the operating conditions considered

Fig 1(a), presents the isometric schematic of the device configuration. The corresponding cross-sectional view in Fig 1(b), taken along the A - A$^{/}$ direction, shows the vertical arrangement of the device layers. Two lateral gates are separated by a well-defined gate gap and are patterned on a silicon body layer on the BOX. The fabricated device has channel width, thickness and gate gap (i.e, the distance between side gates and the channel) of 100 nm.

The devices consist of three regions named as the source extension ($X_I$), channel ($X_{II}$), and drain extension ($X_{III}$). In the fabricated device, source and drain extensions are 2 µm and channel region (the region under the influence of the side gate's electrostatic effect) is fixed at 200 nm. The device active region is Si with source/channel/drain doping concentrations set to $10^{15}$ cm$^{-3}$ matching the nominal fabricated device. Fig. 1(c) provides a scanning electron microscopy (SEM) image of the fabricated device, confirming the accuracy of gate placement and channel definition, and validating the compatibility of the fabrication process with the proposed architecture.

In the TCAD simulations, the effects of geometrical parameters—including channel width, channel thickness, lateral gate gap, and source/drain extension length- were systematically investigated. During the parametric study, the channel width was varied from 100 nm to 30 nm, while the channel thickness ranged from 100 nm to 20 nm. The lateral gate-to-channel air gap was adjusted between 150 nm and 50 nm. In addition, source and drain extension lengths of 500 nm, 1 µm, and 2 µm were considered to evaluate their impact on carrier transport, velocity, and mobility

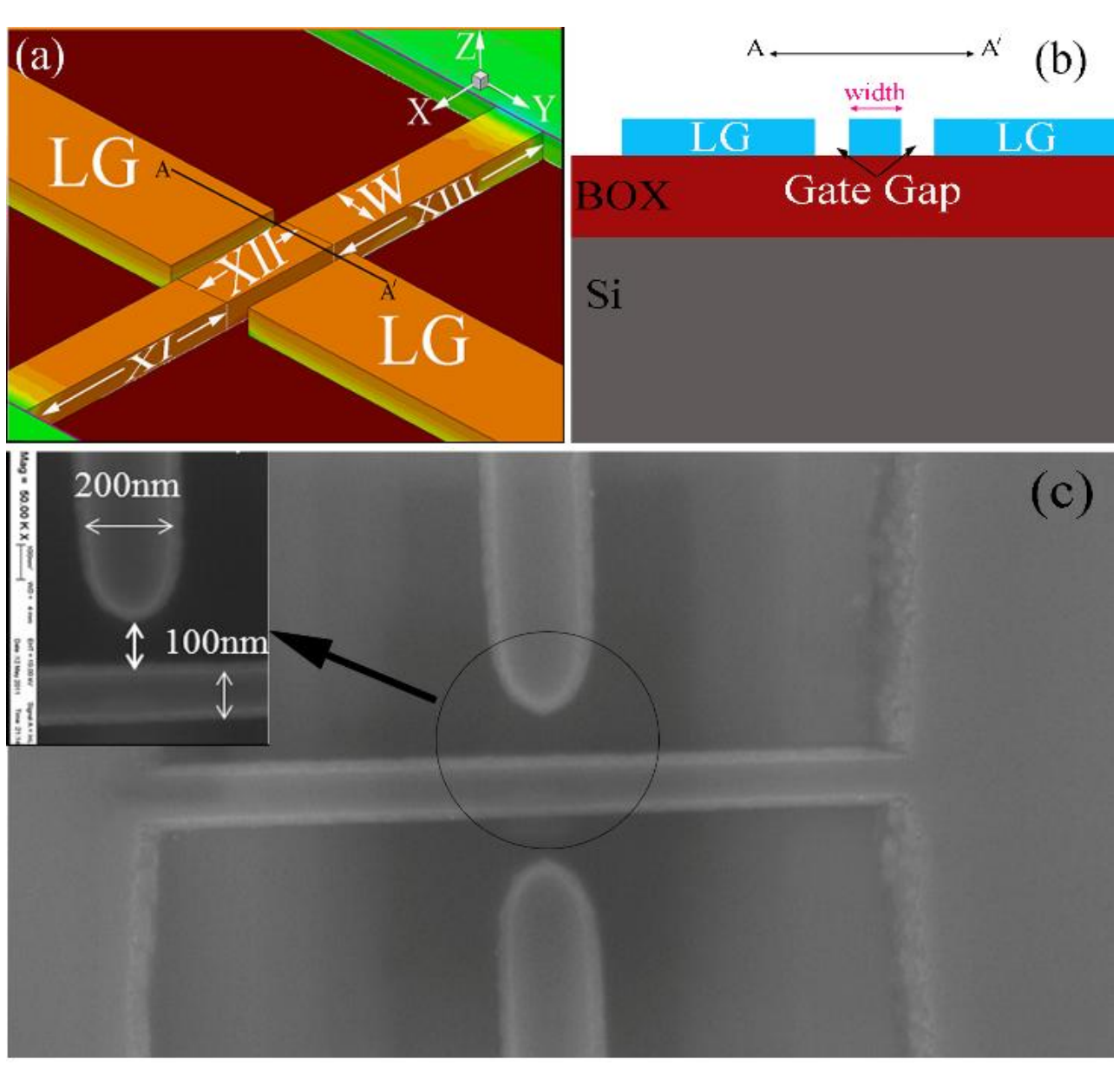

Fig. 1 (a) Isometric schematic of the device, illustrating the configuration of the lateral gates and the silicon channel. The device width and various cross-sectional directions consist of source extension $X_I$, channel region ($X_{II}$) and drain extensions ($X_{III}$) are indicated. (b) Cross-sectional view along A - A′ showing the gate gap, lateral gates (LG), and silicon channel on a BOX layer, representing a typical SOI structure. (c) SEM image of the fabricated device confirming the alignment and geometry of the lateral gates relative to the source and drain regions.

## III. Results and Discussions

The electrical characteristics of the DLGJLT device were measured at room temperature using a semiconductor parameter analyzer (Lakeshore/Desert Cryogenics, Agilent HP 4156C). Fig 2(a) presents the simulated output characteristics under varied gate biases, while Fig 2(b) compares the experimentally measured and simulated transfer characteristics at two drain voltages. The threshold voltage and flat-band voltage are key parameters in this analysis, as they define the practical operating range of the device. The device is in *on-state* at zero gate bias (inset of Fig 2b). The principle of the device charge transmission was discussed in details elsewhere [28].

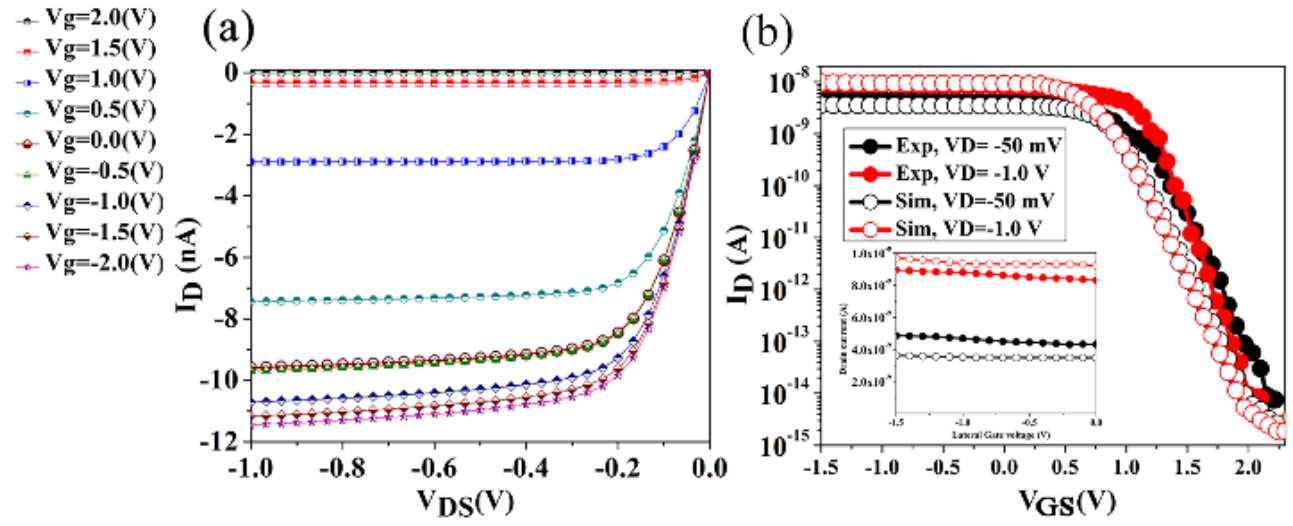

Fig. 2(a) Simulated $I_D$-$V_{DS}$ with $V_{GS}$ varied from -2V to +2V and (b) comparison of the measured and simulation $I_D$ x $V_{GS}$ curves with $V_{DS}$ of -50 mV and -1.0 V. The inset shows the on state region of the device.

The simulation results exhibit good agreement with the experimentally measured data. As shown in Fig 2(b), the device remains in the *on-state* under zero gate bias, whereas applying a sufficient positive voltage to the lateral gates depletes the hole channel and the device into *off-state* (pinch-*off*), demonstrating effective electrostatic control of hole conduction. The transfer characteristics further indicates an approximate on/off ratio of $10^6$ for gate voltages ranging from 0 to +2 V. The observed shift between experimental and simulated transfer characteristics along the $V_{GS}$ axis is attributed to non-ideal interface effects not explicitly included in the TCAD simulations. In particular, fixed charges and interface trap states at the Si/air and Si/$SiO_2$ interfaces, arising from fabrication and surface exposure, can introduce additional depletion in the p-type channel, resulting in a positive shift of the threshold (flat-band) voltage. In the simulations, ideal interface conditions were assumed to focus on transport behavior and mobility trends rather than threshold voltage fitting. Fig 3 shows comparison of experimental and simulated drain conductance ($g_D$) under two different gate voltages. The trend for $g_D$ follows the expected similar trend with MOSFETs [40] and JLTs [41], yet the slope is smaller here which can be explained by low doping concentration of the Si channel hence the current value.

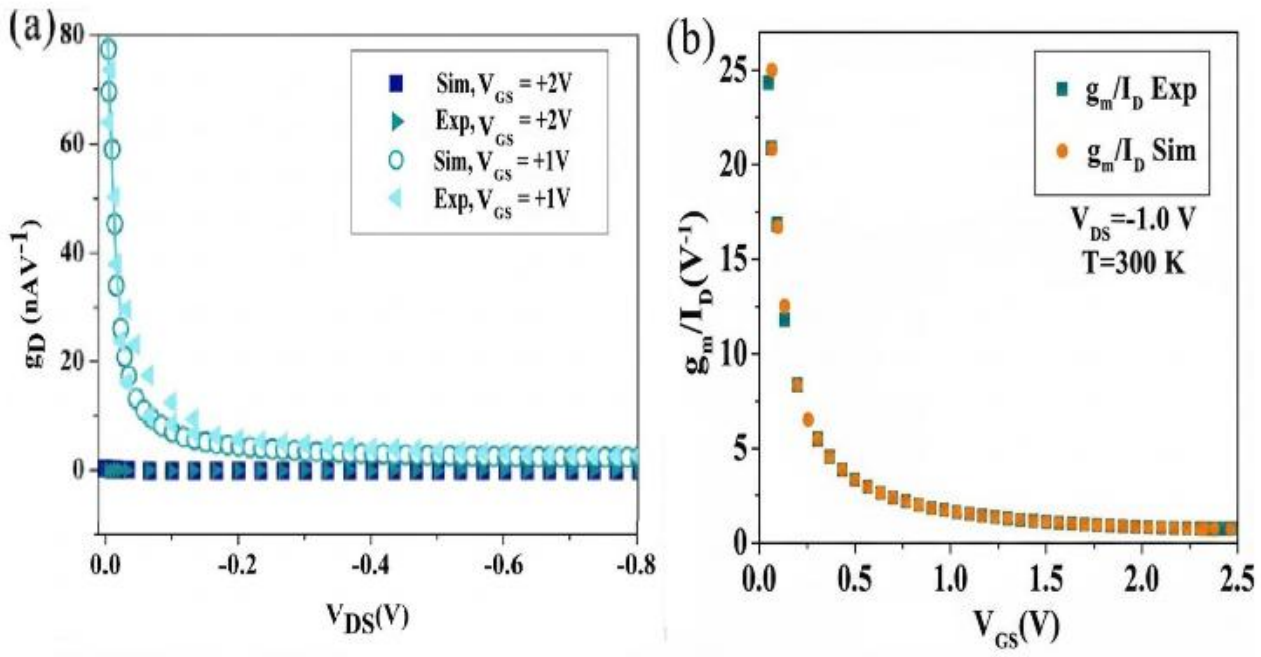

Fig. 3 Drain conductance $g_D$ x $V_{DS}$ and (inset) transconductance $g_m/I_D$ of the fabricated and simulated device.

The inset image in Fig 3, compares experimental and numerical values, when the threshold voltage can be extracted from the gate voltage at which the curve $g_m/I_D$ ($g_m$ = transconductance) drops to half of its maximum value. Conventional MOSFETs employ a gate electrode separated from the channel by an oxide layer, enabling electrostatic control through adjustments in oxide thickness, dielectric material, and gate length [18, 40, 42-44].However, positioning the gate oxide directly above the channel in can generate strain-related impacts, like oxidation-induced strain (OIS) [45] and metal-gate-induced strain [46, 47]. These strain mechanisms can modify the valence-band structure of p-type channels, creating anisotropy in the heavy-hole band. The oxide/semiconductor interface at the top or sides of the channel may introduce additional interface-related scattering and mobility degradation due to atomic-scale irregularities and defects. Limited control and spatial non-uniformity of stress existed at these interfaces can degrade carrier mobility by generating additional roughness and defects.

In standard JLT devices, the high channel doping further introduces multiple scattering mechanisms, e.g. impurity, acoustic-phonon, optical-phonon, and polar-optical-phonon scattering [48, 49]. Interface-related mobility degradation is commonly associated with interface irregularities, wavefunction penetration, and confinement-related effects [50]. In contrast, the absence of a vertical gate interface and the use of laterally positioned gates combined with a lightly doped channel in the DLGJLT architecture can mitigate several of these unwanted scattering mechanisms.

Under typical p-type MOSFET operation, hole transport is concentrated near the low-energy region around the Γ-point. Fig 4(a) illustrates the 100-meV iso-energy contour of the heavy-hole (HH) band under zero stress, consisting of 12 wings (eight out-of-plane wings O (1–8) and four in-plane wings I (1–4)). Mobility is governed not only by states at the Γ-point but also by how these wings respond to mechanical stress. Biaxial tensile stress modifies the energy of in-plane wings, shifting hole occupation toward the off-plane wings and reshaping their contours [47]. Under high gate fields, gate-oxide-induced tensile strain further couples the off-plane wings with the $k_z$ quantization direction, lowering their energy and increasing occupancy of hole states with reduced effective mass ($m_{eff}$). Within the parabolic-band approximation, carrier velocity scales inversely with $m_{eff,}$ hence, reduced effective mass leads to mobility enhancement. Because the DLGJLT structure eliminates gate-induced mechanical stress and confines holes predominantly within the in-plane wings, improved transport characteristics are expected compared to devices employing top-gate, tri-gate, or GAA configurations. This configuration minimizes mismatch effects and reduces interface-related scattering, except for the interface with the buried oxide (BOX). The difference in transport behavior can be further understood by examining the hole-band wing shapes in Fig. 4b, which displays 25-meV iso-energy contours in the (001) plane for the lowest-energy hole band under zero stress [34]. I1 and I3 have more curvature in [110] direction compare to I2, I4. Therefore, carriers are expected to be located in I1, I3 with lighter $m_{eff}$ than those on I2, I4. If we align the channel in [110] we would have more carriers mobility along the channel [51]. In fabrication process, [52-54] we kept the channel direction in [110] direction to benefit from this advantage, when the channel aligned the [110] crystal direction. Although the effect of the strain on the SOI wafer to have a mobility improvement for JLTs has been reported in literature [6], but while it remains to be quantitatively determined, we believe that the lack of the mechanical stress of the gate on the channel would improve the mobility of carriers according to the hole energy distribution [47].

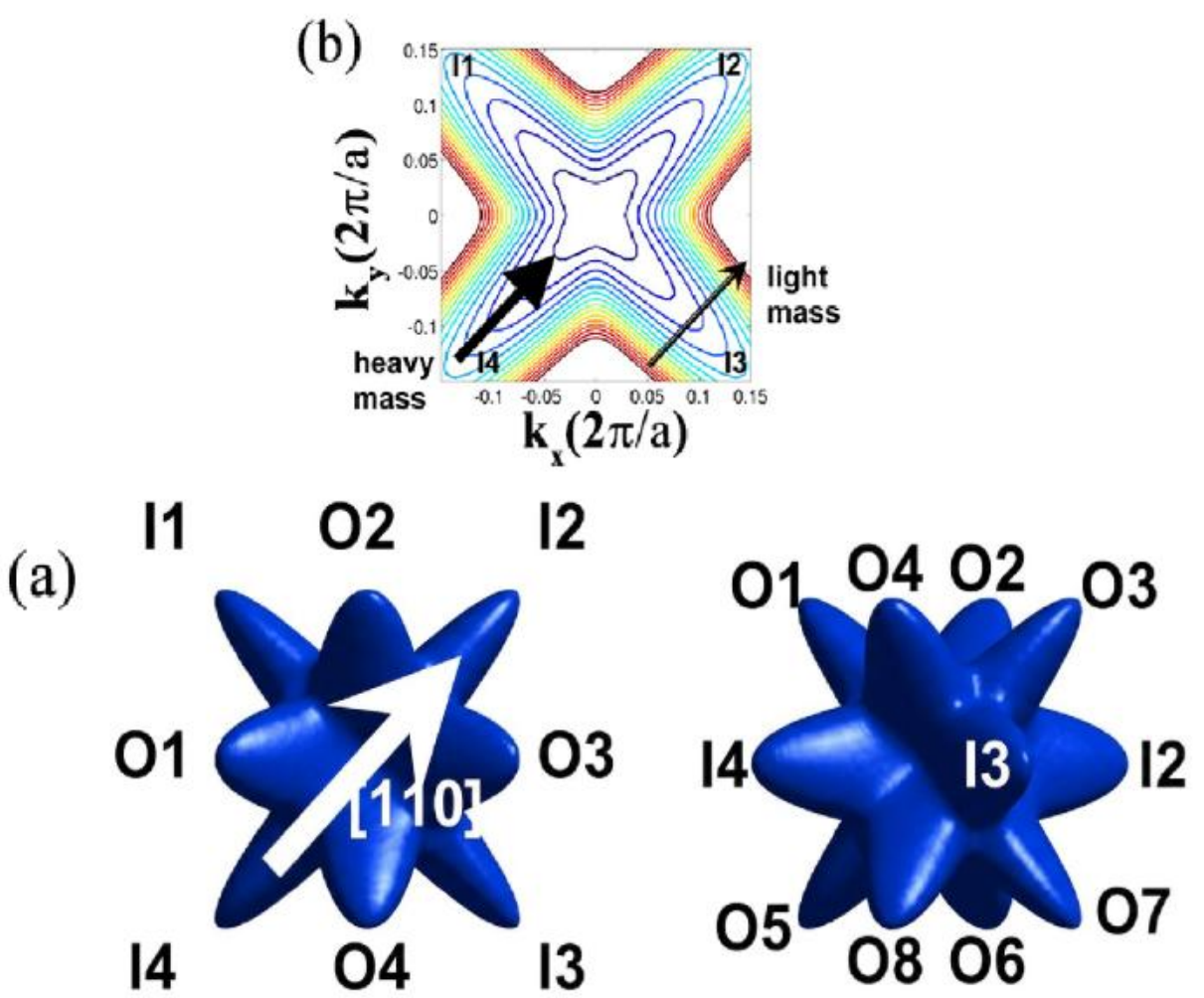

Fig. 4 (a) Top and side view of bulk heavy hole (HH) Iso-surfaces under no stress and (b) iso-energy contour separated by 25 meV in the (001) plane for the lowest energy hole band under no stress condition (modified from [36]).

In Fig 5(a) hole mobility along a horizontal cut line at the center of the channel (Z= 50 nm) along the X-axis at gate voltages ($V_{GS}$) of +0.5, +1.0, +1.5, +2.0 V and in Fig 5(b) a vertical cut at y= 0 at $V_{DS}$= -1.0 V, $V_{GS}$= +0.5 V are shown. In the DLGJLTs since the conduction path is located near the center of the channel in the flat band voltage ($V_{FB}$), the carriers exhibit bulk properties and traverse the channel with the mobility value comparable to the bulk mobility. In Fig 5(b) we have applied a small positive gate voltage to show the effect of side gates in the beginning of the pinch-off on the carriers' mobility. By increasing the positive applying bias on the gates, the hole mobility significantly decreases. This can be explained by the increased carrier scattering resulted from the lateral gate's vertical electric field. This field repels the holes to the middle of the channel and this redistribution causes higher transverse electric field, increasing phonon scattering and reducing mobility. The accumulation of the carriers in the middle of the channel enhancing carrier-carrier scattering and limiting mobility.

The lateral gate voltage modifies the valence-band profile through electrostatic band bending, redistributing holes toward higher-energy transport states within the valence band. This redistribution increases the energy-averaged effective transport mass and enhances field-dependent scattering, leading to reduced hole mobility. The effect of band bending on carrier transport is captured self-consistently through the hydrodynamic transport equations together with the doping-dependent Masetti mobility model, which account for carrier heating, non-equilibrium transport, velocity saturation, and field-enhanced scattering under strong lateral electric fields. Therefore, the observed mobility degradation originates primarily from electrostatic confinement and high-field transport effects rather than explicit quantum confinement.

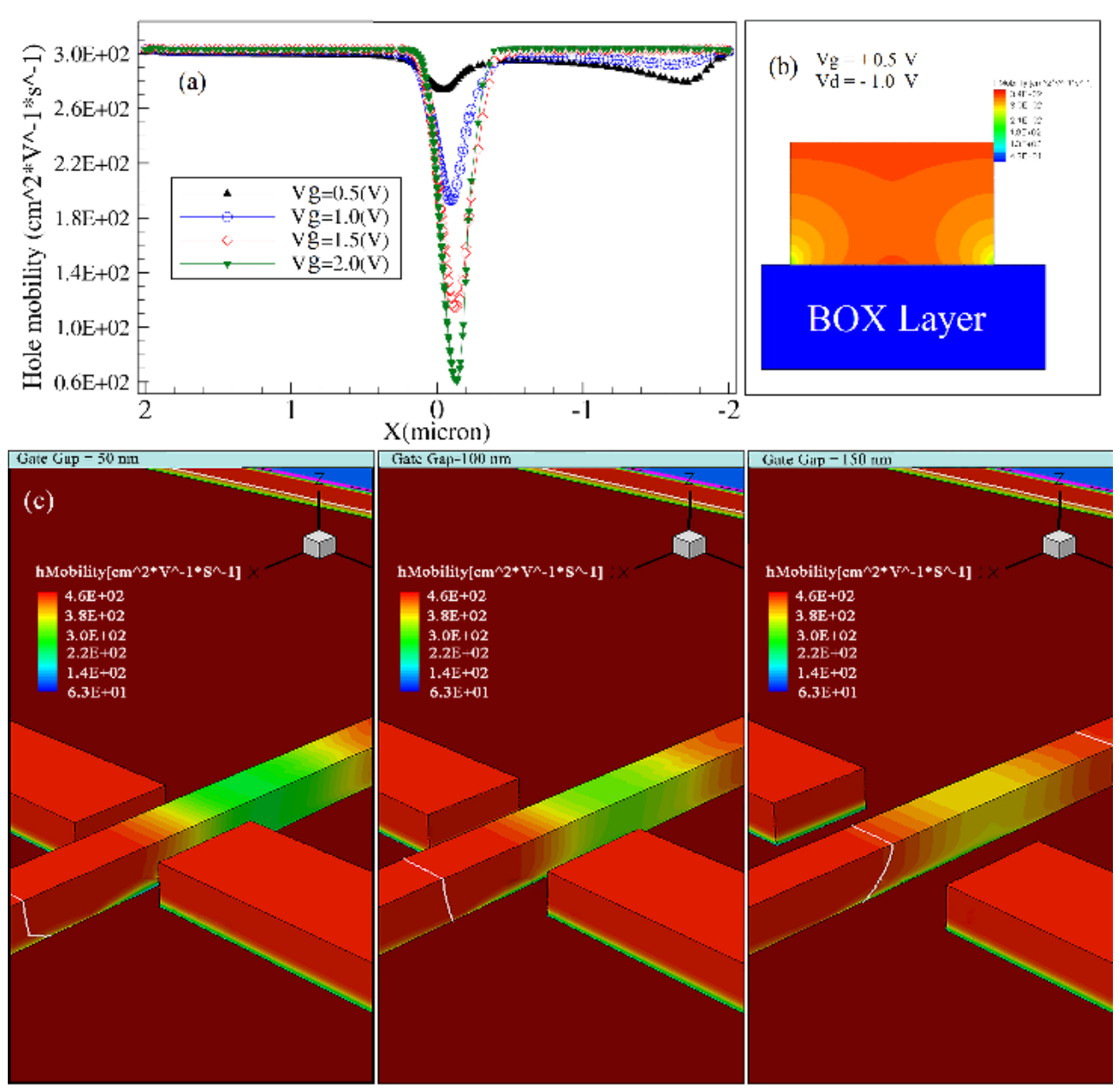

Fig 5. (a) Hole mobility along a horizontal cut line at the center of the channel (Z= 50 nm) along the X-axis at gate voltage of +0.5, +1.0 , +1.5, +2.0 V and (b) a vertical cut at y= 0 at $V_{DS}$= -1.0 V, $V_{GS}$= +0.5 V. (c) an isometric view of hole mobility for devices with different gate gap of 50, 100, and 150 nm in the pinch off region, $V_{GS}$ = +2.0 V and $V_{DS}$ = −1.0 V.

In Fig 5(c), the effects of different gate gap of 50, 100, and 150 nm in the pinch-off region, $V_{GS}$ = +2.0 V and $V_{DS}$ = −1.0 V are presented. As the gate-to-channel gap is reduced in the simulations, the lateral confinement becomes stronger, an effect analogous to applying a higher gate voltage. This increased confinement modifies the valence-band structure, resulting in a higher hole $m_{eff}$ and consequently lower mobility. The influence of channel-width scaling on hole transport was further investigated by examining the variations in hole velocity and mobility as a function of the electric field along the channel.

As illustrated in Fig. 6, decreasing the channel width from 100 nm to 60 nm and then to 30 nm leads to pronounced changes in carrier behavior within the pinch-off region. The primary mechanism responsible for this trend is the enhanced lateral confinement experienced by holes in narrower channels. With decreasing channel width, spatial quantization intensifies, inducing substantial alterations in the valence-band characteristics, especially the coupling between heavy-hole and light-hole bands. This enhanced confinement increases the effective mass of the holes and therefore degrades their phonon- and impurity-limited mobility. Although the lateral-gates electrostatics remain comparable for all devices, the intrinsic transport degrades as the channel width is scaled down. Narrower channels allow the lateral-gates field lines to penetrate more strongly into the silicon, producing a higher and more non-uniform electric field, especially in the pinch-off region. For a fixed gate bias, this intensified field increases the carrier drift velocity, which explains the higher velocity observed in the 30-nm device.

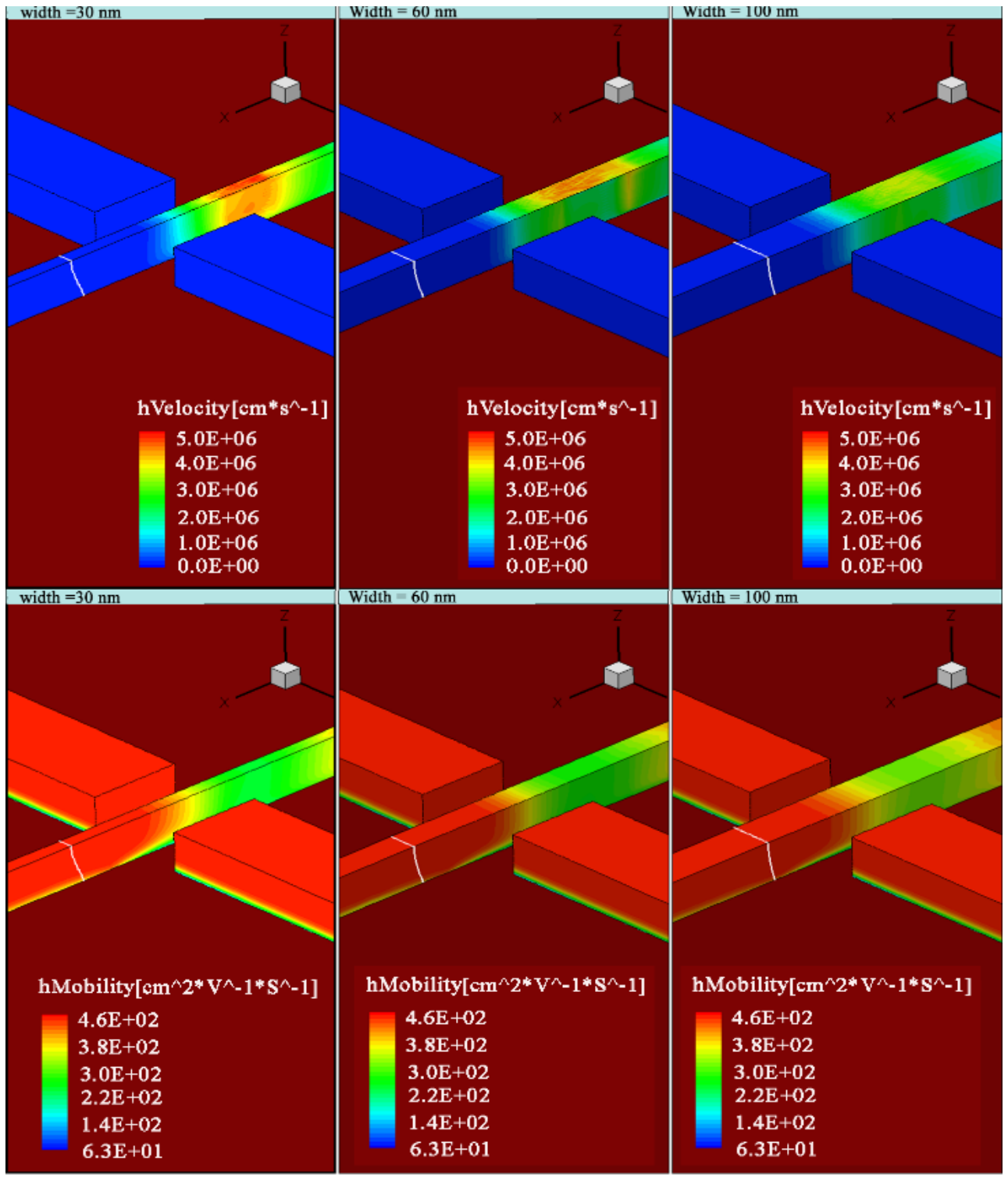

Fig. 6 Side view of carriers' velocity and mobility for devices with different width of 30, 60, and 100 nm in the pinch off region, $V_{GS}$ = +2.0 V and $V_{DS}$ = −1.0 V

However, the extracted mobility decreases because mobility is limited by momentum-relaxing scattering rather than instantaneous drift velocity. The stronger fields enhance carrier scattering and high-field transport effects, reducing the effective mobility, while at sufficiently high fields the carriers approach velocity saturation. Overall, the combined effects of width-induced quantum confinement and field-driven transport modulation govern the observed trends. Wider channels exhibit weaker confinement and smoother field profiles, yielding higher hole mobility but lower drift velocity. Conversely, narrow channels impose strong confinement and sharper electric-field gradients, which increase drift velocity while reducing mobility. The resulting mobility–velocity trade-off reflects the interplay between band-structure modification, enhanced scattering in confined geometries, and field intensification in the lateral-gate configuration. These findings demonstrate that channel-width scaling provides an additional geometric handle for tuning carrier transport in DLGJLT devices.

Fig 7(a) shows the magnitude of the net electric field (|E|), obtained from the vector norm of the electric field along the channel of the DLGJLT device for body thicknesses of 20 nm, 40 nm, and 100 nm at $V_{GS}$ = +2.0 V and $V_{DS}$ = -1.0 V. The results reveal a distinct transition in the electric-field behavior as the channel thickness decreases. For the device with 100 nm thickness, the electric field shows an increase at the center, forming a peak, while as the thickness of the device decreases (40 nm and 20 nm), the electric field exhibits a pronounced dip at the channel center. This difference arises from the fundamental change in the electrostatics and depletion behavior of device when the silicon body thickness varies.

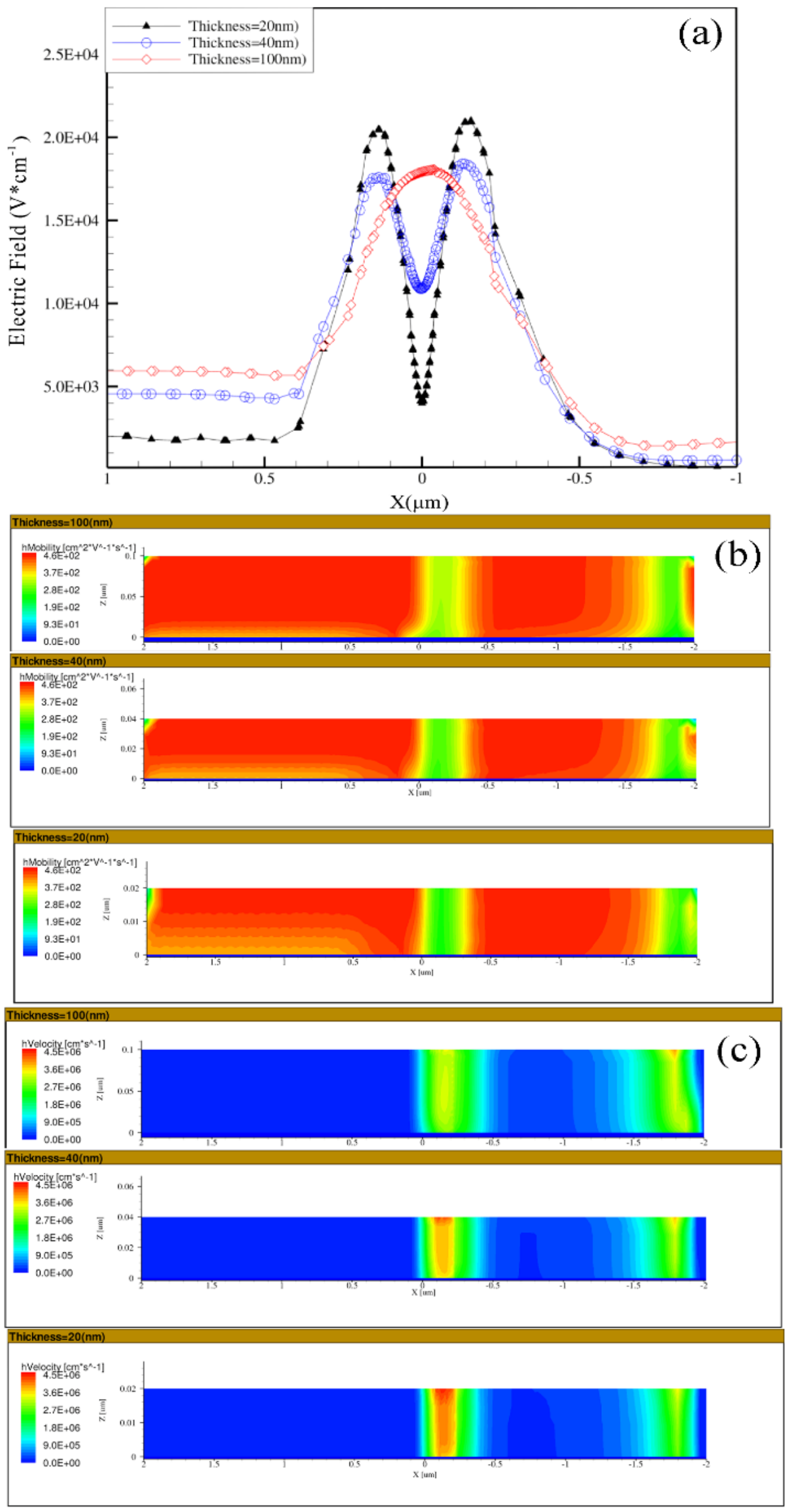

Fig. 7 (a) net electric field (|E|) variation along a horizontal cut line at the center of the channel along the X-axis (Y=0 and Z=50, 20, and 10 nm) (b) carriers' mobility and (c) Side view of carriers' velocity along a vertical cut at Y = 0 for devices with different thickness of 20, 40, and 100 nm. $V_{GS}$ = +2.0 V and $V_{DS}$ = -1.0 V.

In device with 100 nm thickness, the gate electric field cannot fully penetrate the body. The depletion regions induced by the double lateral gates are confined primarily to the outer regions near the gate interfaces, leaving a neutral, un-depleted core at the center of the channel. This transition corresponds to the device entering a partially depleted regime, where the gate loses strong electrostatic control over the central region of the silicon film. In this case, the electric field components cannot be cancelled at the center; instead, the central region contains a higher concentration of ionized dopants and remains more conductive. Since the electric field in a junctionless device is directly related to the space-charge density (via Gauss's law, $E=\rho/\varepsilon$, the presence of this neutral or lightly depleted core results in a higher local charge density and therefore a higher electric field magnitude in the mid-channel region. This explains the upward shift and formation of a central peak in the electric-field profile for the 100 nm device. In contrast, as the channel thickness decreases (40–20 nm), the electrostatic influence of the lateral gates extends across the entire silicon thickness. As a result, the channel becomes fully depleted under gate bias, and the depletion regions originating from the lateral gates overlap at the center of the body. In this fully depleted regime, the mobile carrier concentration is significantly reduced throughout the film. The electric field generated by the two gates is also symmetrical, and the field components propagating inward from each side tend to oppose and partially cancel at the center of the channel. This cancellation leads to a lower electric field magnitude in the central region, producing the observed valley in the electric-field profile for 40 nm and 20 nm thicknesses. This behavior is a characteristic signature of strong electrostatic gate control and full-body depletion in thin junctionless structures.

The distributions of majority carriers' mobility, and velocity in the device are directly governed by the electrostatics associated with variations in silicon body thickness. As the thickness decreases from 100 nm to 40 nm and 20 nm (Fig 7(b) and 7(c)), the device transitions from a partially depleted regime to a fully depleted regime, and this transition explains the distinct behaviors observed in the electric-field and carrier-transport profiles. For the 100-nm device, based on electrostatic behavior, not fully depleted channel, allowing a neutral conductive core to persist at the center of the channel. This core contains a higher free-hole concentration and thus experiences strong impurity scattering, which reduces the intrinsic low-field mobility. However, the greater distance from the interfaces mitigates interface-related scattering effects, partially offsetting the impurity-scattering penalty. More importantly, the presence of the not fully depleted core alters the lateral electric-field distribution: the field no longer exhibits a dip at the center but instead reaches a peak. Since the depletion regions on either side do not overlap, the lateral field components do not cancel, and the resulting nonuniform potential distribution generates a strong field maximum at the mid-plane. This elevated electric field increases the drift velocity in the center, even if the mobility is moderately reduced by scattering or field dependence. For the 40 nm and 20 nm structures, the lateral gates exert strong electrostatic control across the entire silicon film, such that the depletion regions extending from each gate merge at the channel center. The merger reduces the free-hole concentration and simultaneously produces a noticeable reduction in the lateral electric field at the mid-plane. In these thin-body devices, although the low carrier density diminishes carrier–carrier and impurity scattering, the carriers that remain are positioned closer to the top and bottom interfaces, where interface-related scattering effects are expected to become more significant. As a result, the mobility near the interfaces is significantly degraded, whereas the interior of the film exhibits moderately higher mobility, consistent with the simulation maps. Because the lateral electric field at the center is small, the drift velocity also reaches a minimum in this region, despite any moderately favorable mobility.

Consequently, current conduction tends to localize near the edges of the depletion boundaries—where the electric field is higher-rather than uniformly across the channel thickness. Overall, the comparison across the three thickness values demonstrates that the conduction mechanism in DLGJLT is highly thickness-dependent. Thin films provide strong gate control and full depletion but suffer from reduced mobility due to interface-related scattering and exhibit lower drift velocity at the channel center due to the suppressed lateral field. In contrast, thick films provide higher drift velocity and enhanced current drive due to the presence of a neutral core and increased central electric field, but they sacrifice electrostatic integrity and become more susceptible to leakage and degraded short-channel behavior. These results confirm that the onset of neutral-core formation marks a critical transition point in junctionless device operation and that the interplay between electrostatic depletion, mobility degradation mechanisms, and field-dependent transport determines the overall carrier behavior observed in simulation. Building upon the earlier analysis of channel width and thickness scaling, where lateral and vertical confinement and electrostatic penetration depth were shown to shape the transverse field distribution and hole transport, the investigation was next extended to assess the impact of source and drain extension length ($X_I$ and $X_{III}$).

Unlike width and thickness variation, which primarily affects vertical band modulation and the strength of gate-induced confinement, the source and drain extension lengths directly modify the longitudinal transport path while leaving the lateral-gate-controlled electrostatic environment largely unchanged. Even though the lateral gates and the air gap maintain a nearly identical electric-field magnitude and spatial trend inside the channel for different source and drain extension, subtle but physically meaningful differences arise in carrier transport and recombination (Figs 8(a) and 8(b)). While the field uniformity ensures that the electrostatic driving force remains comparable, the effective mobility decreases as channel length increases due to the longer carrier transit path and the cumulative influence of phonon, impurity, and surface-related scattering mechanisms. In devices with shorter source/drain extension length, carriers experience fewer scattering events before reaching the drain, resulting in a higher average drift velocity despite similar electric field. In contrast, longer devices introduce additional scattering opportunities along the extended transport path, lowering the effective mobility even when the steady-state field distribution is nearly unchanged. The longer transit time also directly contributes to an increase in Shockley–Read–Hall (SRH) recombination. Because carriers stay within the channel for a longer duration, the probability of being captured by interface and bulk traps increases, and the rate of SRH generation–recombination rises accordingly. Thus, the observed growth in SRH recombination with increasing channel length (Fig 8 (c)) arises not from substantial electric-field modifications but from the intrinsic time-dependent likelihood of trap interactions over longer transport distances.

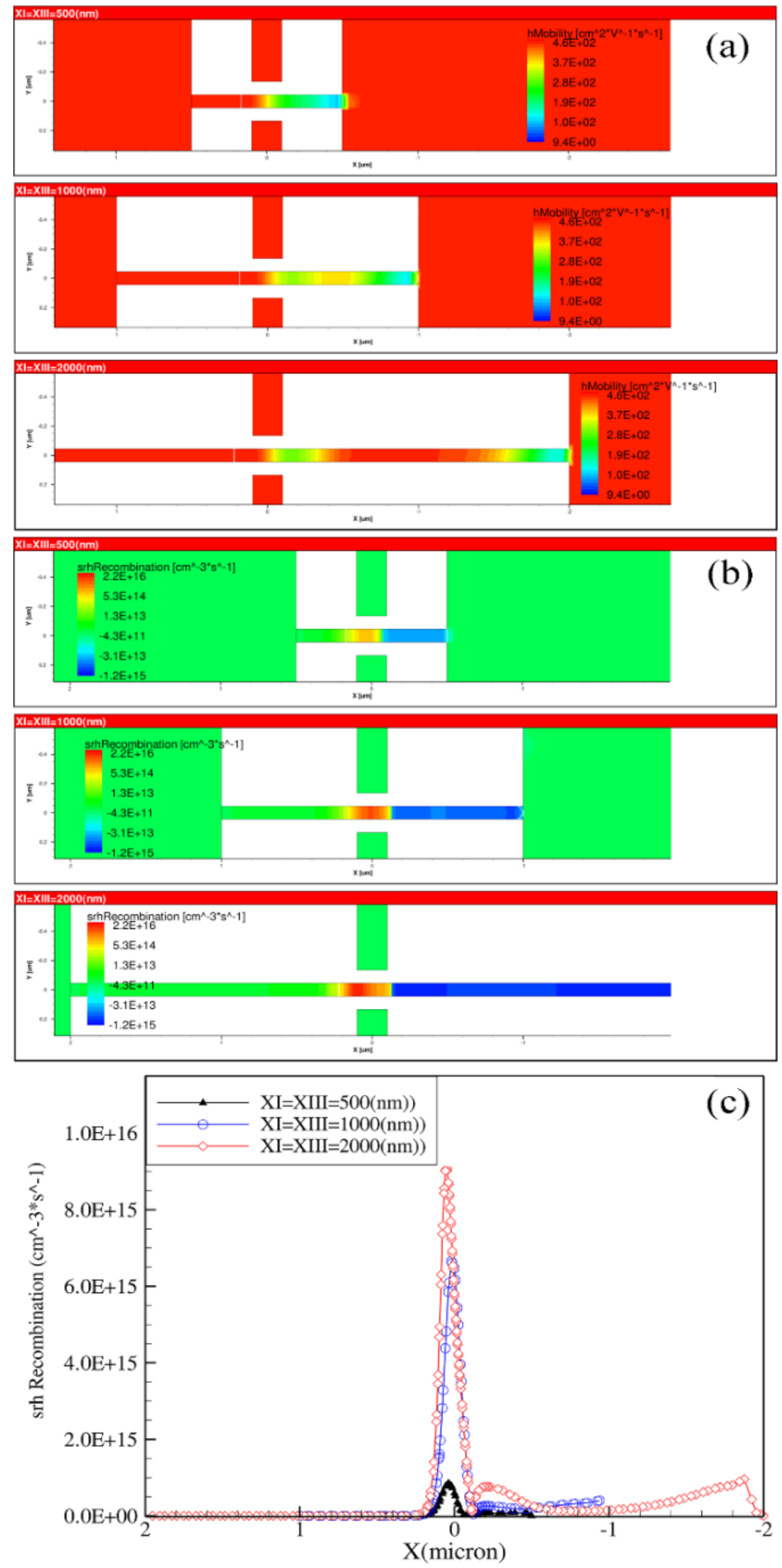

Fig 8. (a) Hole mobility and (b) srh Recombination along a horizontal cut line at the center of the channel (Z= 50 nm) along the X-axis at gate voltage of +2.0 V and $V_{DS}$= -1.0 V (c) SRH recombination along a horizontal cut line at the center of the channel (Z= 50 nm) along the X-axis at $V_g$=+2.0 V and $V_{DS}$=-1.0 V.

These results demonstrate that even under nearly uniform electrostatic conditions provided by the lateral-gate configuration, source and drain extension scaling strongly influences mobility degradation and recombination behavior in DLGJLT architectures. The high-mobility pocket observed near the drain extension in the $X_I$=$X_{III}$=2 μm device can be associated with reduced interface-related scattering and improved impurity screening at that location, combined with the numerical definition of local mobility; vertical redistribution of carriers away from the Si/oxide interface and a locally reduced perpendicular field lead to higher effective mobility despite unchanged lateral gate field.

## IV. Conclusions

We focused on the carrier transport characteristics of the DLGJLT. Experimental and simulated device were compared. The device operates normally-on and achieves full channel depletion under positive lateral-gate bias, with pinch-off behavior consistent with junctionless architecture. Further scaling studies showed that channel geometry strongly influences transport despite the nearly uniform electrostatics produced by the lateral gates. Reducing the channel width enhances lateral field coupling and increases carrier velocity but degrades mobility due to stronger confinement and increased effective mass. Channel thickness strongly influences the depletion regime: thick channels retain a partially depleted conductive core with higher velocity but increased impurity scattering, whereas thinner channels achieve full depletion with more uniform electrostatic control at the cost of reduced mobility. Source/drain extension scaling mainly affects transport through changes in carrier transit length and recombination, with longer extensions increasing scattering and SRH recombination while leaving the internal field distribution largely unchanged.

These results establish practical design guidelines for DLGJLTs: moderate channel widths and thicknesses offer the best compromise between mobility and velocity, while shorter source/drain extensions minimize recombination losses. The presented insights provide a framework for geometry optimization of lateral-gate junctionless transistors in nanoscale and low-power applications.

## References

[1] P. Rajeswari, A. Gobinath, N. Suresh Kumar, and M. Anandan, "Nanoscale Field-Effect Transistors (FETs) in RF Applications," *Field Effect Transistors,* pp. 443-455, 2025.

[2] B. Khan, A. Mukherjee, Y. M. Georgiev, J.-P. Colinge, S. Ghosh, and S. Das, "Observation of room-temperature gate-tunable quantum confinement effect in a photodoped junctionless MOSFET," *Physical Review B,* vol. 112, no. 4, p. 045423, 2025.

[3] J.-P. Colinge *et al.*, "Junctionless nanowire transistor (JNT): Properties and design guidelines," *Solid-State Electronics,* vol. 65, pp. 33-37, 2011.

[4] J. P. Colinge *et al.*, "Nanowire transistors without junctions," *Nature Nanotechnology,* vol. 5, no. 3, pp. 225-229, 2010.

[5] Y. Siddiqui, I. U. Khan, and S. H. Saeed, "A Comprehensive Review Of Junctionless Transistors," in *2024 Second International Conference Computational and Characterization Techniques in Engineering & Sciences (IC3TES)*, 2024: IEEE, pp. 1-4.

[6] J.-P. Raskin *et al.*, "Mobility improvement in nanowire junctionless transistors by uniaxial strain," *Applied Physics Letters,* vol. 97, no. 4, 2010.

[7] V. B. Sreenivasulu, M. Prasad, E. Deepthi, A. S. Kumar, and S. S. Mangalampalli, "Analysis of novel core-shell junctionless nanosheet FET for CMOS logic applications," *IEEE Access,* vol. 12, pp. 144479-144488, 2024.

[8] D. Sels, B. Sorée, and G. Groeseneken, "Quantum ballistic transport in the junctionless nanowire pinch-off field effect transistor," *Journal of Computational Electronics,* vol. 10, no. 1, pp. 216-221, 2011/06/01 2011, doi: 10.1007/s10825-011-0350-2.

[9] C. W. Lee *et al.*, "High-Temperature Performance of Silicon Junctionless MOSFETs," *IEEE Transactions on Electron Devices,* vol. 57, no. 3, pp. 620-625, 2010, doi: 10.1109/TED.2009.2039093.

[10] S. S. Mohanty, S. Mishra, and G. P. Mishra, "Impact of Sensing Performance on Junction Less Stack Gate All-Around Metal Oxide Semiconductor Field Effect Transistor for Biosensing Application," in *2025 Devices for Integrated Circuit (DevIC)*, 2025: IEEE, pp. 474-478.

[11] Y. Yan *et al.*, "A planar core-shell junctionless transistor compatible with FD-SOI Technology," *Solid-State Electronics,* vol. 225, p. 109079, 2025.

[12] R. Sharma and P. Karupannan, "Enhancing Performance of SiGe/InAs Junctionless TFET," in *2024 3rd International Conference for Innovation in Technology (INOCON)*, 2024: IEEE, pp. 1-4.

[13] R. R. Prates, S. Barraud, M. Cassé, M. Vinet, O. Faynot, and M. A. Pavanello, "Comprehensive evaluation of junctionless and inversion-mode nanowire MOSFETs performance at high temperatures," *IEEE Journal of the Electron Devices Society,* vol. 12, pp. 682-691, 2024.

[14] A. Mohanty, M. A. Ahmad, P. Kumar, and R. Kumar, "Performance analysis and design comparison of junctionless TFET: A review study," *Silicon,* vol. 16, no. 18, pp. 6305-6312, 2024.

[15] K. Kumar, A. Kumar, V. Kumar, A. Jain, and S. C. Sharma, "Band gap and gate dielectric engineered novel Si0. 9Ge0. 1/InAs junctionless TFET for RFIC applications," *Engineering Research Express,* vol. 6, no. 3, p. 035340, 2024.

[16] W.-T. Chang, C.-C. Yang, Y.-T. Ho, C.-L. Wang, and W.-I. Shen, "Electrical Stress on the CMOS Inverters Made by Junctionless Gate-All-Around Transistors," *IEEE Transactions on Electron Devices,* vol. 71, no. 5, pp. 2863-2868, 2024.

[17] S. Bhattacharya, S. L. Tripathi, and G. Nayana, "Design transmission gates using double-gate junctionless TFETs," *Silicon,* vol. 16, no. 8, pp. 3359-3372, 2024.

[18] O. Durante *et al.*, "Subthreshold current suppression in ReS2 nanosheet-based field-effect transistors at high temperatures," *ACS Applied Nano Materials,* vol. 6, no. 23, pp. 21663-21670, 2023.

[19] P. Dasika *et al.*, "Contact-barrier free, high mobility, dual-gated junctionless transistor using tellurium nanowire," *Advanced Functional Materials,* vol. 31, no. 13, p. 2006278, 2021.

[20] A. Nowbahari, A. Roy, and L. Marchetti, "Junctionless transistors: State-of-the-art," *Electronics,* vol. 9, no. 7, p. 1174, 2020.

[21] A. Konar *et al.*, "Carrier transport in high mobility InAs nanowire junctionless transistors," *Nano letters,* vol. 15, no. 3, pp. 1684-1690, 2015.

[22] A. Dehzangi *et al.*, "Numerical investigation and comparison with experimental characterisation of side gate p-type junctionless silicon transistor in pinch-off state," *Micro & Nano Letters,* vol. 7, no. 9, pp. 981-985, 2012.

[23] A. Dehzangi *et al.*, "Study the Characteristic of P-type Junction-less Side Gate Silicon Nanowire Transistor Fabricated by AFM Lithography," *American Journal of Applied Science,* vol. 8, no. 9, pp. 872-877, 2011.

[24] M. Kumar *et al.*, "A comprehensive study of junctionless TFETs as a low power device," *Analog Integrated Circuits and Signal Processing,* vol. 126, no. 1, p. 7, 2026.

[25] A. K. Panigrahy *et al.*, "Analysis of GAA junction less NS FET towards analog and RF applications at 30 nm regime," *IEEE Open Journal of Nanotechnology,* vol. 5, pp. 1-8, 2024.

[26] F. Larki *et al.*, "Electronic transport properties of junctionless lateral gate silicon nanowire transistor fabricated by atomic force microscope nanolithography," *Microelectronics and Solid State Electronics,* vol. 1, no. 1, pp. 15-20, 2012.

[27] F. Larki *et al.*, "Pinch-off mechanism in double-lateral-gate junctionless transistors fabricated by scanning probe microscope based lithography," *Beilstein journal of nanotechnology,* vol. 3, no. 1, pp. 817-823, 2012.

[28] A. Dehzangi *et al.*, "Electrical property comparison and charge transmission in p-type double gate and single gate junctionless accumulation transistor fabricated by AFM nanolithography," *Nanoscale research letters,* vol. 7, no. 1, p. 381, 2012.

[29] A. Dehzangi *et al.*, "Impact of KOH etching on nanostructure fabricated by local anodic oxidation method," *International Journal of Electrochemical Science,* vol. 8, no. 6, pp. 8084-8096, 2013.

[30] P. Campbell, E. Snow, and P. McMarr, "Fabrication of nanometer scale side gated silicon field effect transistors with an atomic force microscope," *Applied Physics Letters,* vol. 66, p. 1388, 1995.

[31] Y. Mo, W. Zhao, D. Huang, F. Zhao, and M. Bai, "Nanotribological properties of precision-controlled regular nanotexture on H-passivated Si

surface by current-induced local anodic oxidation," *Ultramicroscopy,* vol. 109, no. 3, pp. 247-252, 2009.
[32] F. Marchi, V. Bouchiat, H. Dallaporta, V. Safarov, D. Tonneau, and P. Doppelt, "Growth of silicon oxide on hydrogenated silicon during lithography with an atomic force microscope," *Journal of Vacuum Science & Technology B: Microelectronics and Nanometer Structures Processing, Measurement, and Phenomena,* vol. 16, no. 6, pp. 2952-2956, 1998.
[33] A. Dehzangi *et al.*, "Atomic force microscope base nanolithography for reproducible micro and nanofabrication," in *2014 IEEE International Conference on Semiconductor Electronics (ICSE2014)*, 2014: IEEE, pp. 408-411.
[34] A. Dehzangi *et al.*, "Fabrication of p-type Double gate and Single gate Junctionless silicon nanowire transistor by Atomic Force Microscopy Nanolithography." Nano Hybrids 3 (2013): 93-113..
[35] J. Martinez, R. V. Martinez, and R. Garcia, "Silicon nanowire transistors with a channel width of 4 nm fabricated by atomic force microscope nanolithography," *Nano letters,* vol. 8, no. 11, pp. 3636-3639, 2008.
[36] A. Dehzangi *et al.*, "Impact of parameter variation in fabrication of nanostructure by atomic force microscopy nanolithography," *PloS one,* vol. 8, no. 6, p. e65409, 2013.
[37] *Sentaurus Process: 1D/2D/3D Process Simulator*. (2010). Synopsys TCAD Documentation, Mountain View, CA, USA.
[38] F. Larki and A. Dehzangi, "View of Charge Carrier Mobility in Lateral Gates Junction-Less Transistor," in *2025 IEEE 25th International Conference on Nanotechnology (NANO)*, 2025: IEEE, pp. 46-51.
[39] F. Larki *et al.*, "Effect of Geometric Parameters on the Performance of P-Type Junctionless Lateral Gate Transistors," *PLOS ONE,* vol. 9, no. 4, p. e95182, 2014, doi: 10.1371/journal.pone.0095182.
[40] A. Nourbakhsh *et al.*, "MoS2 field-effect transistor with sub-10 nm channel length," *Nano letters,* vol. 16, no. 12, pp. 7798-7806, 2016.
[41] J. P. Duarte, S. J. Choi, and Y. K. Choi, "A Full-Range Drain Current Model for Double-Gate Junctionless Transistors," *Electron Devices, IEEE Transactions on,* no. 99, pp. 1-7, 2011.
[42] S. De Stefano *et al.*, "Multilayer MoS 2 Schottky Barrier Field Effect Transistor," *IEEE Open Journal of Nanotechnology,* 2025.
[43] A. Di Bartolomeo *et al.*, "Temperature-dependent photoconductivity in two-dimensional MoS2 transistors," *Materials Today Nano,* vol. 24, p. 100382, 2023.
[44] X. Jing *et al.*, "Engineering field effect transistors with 2D semiconducting channels: Status and prospects," *Advanced Functional Materials,* vol. 30, no. 18, p. 1901971, 2020.
[45] P. Hashemi, L. Gomez, and J. L. Hoyt, "Gate-all-around n-MOSFETs with uniaxial tensile strain-induced performance enhancement scalable to sub-10-nm nanowire diameter," *Electron Device Letters, IEEE,* vol. 30, no. 4, pp. 401-403, 2009.
[46] M. Najmzadeh, D. Bouvet, P. Dobrosz, S. Olsen, and A. M. Ionescu, "Investigation of oxidation-induced strain in a top-down Si nanowire platform," *Microelectronic engineering,* vol. 86, no. 7-9, pp. 1961-1964, 2009.
[47] E. X. Wang *et al.*, "Physics of hole transport in strained silicon MOSFET inversion layers," *Electron Devices, IEEE Transactions on,* vol. 53, no. 8, pp. 1840-1851, 2006.
[48] D. Madadi and A. A. Orouji, "Scattering mechanisms in β-Ga2O3 junctionless SOI MOSFET: investigation of electron mobility and short channel effects," *Materials Today Communications,* vol. 26, p. 102044, 2021.
[49] D.-Y. Jeon, S. J. Park, M. Mouis, S. Barraud, G.-T. Kim, and G. Ghibaudo, "Low-temperature operation of junctionless nanowire transistors: Less surface roughness scattering effects and dominant scattering mechanisms," *Applied Physics Letters,* vol. 105, no. 26, 2014.
[50] F. Chen, K. Wei, E. Wei, and J. Z. Huang, "Hole mobility model for Si double-gate junctionless transistors," in *2017 IEEE Electrical Design of Advanced Packaging and Systems Symposium (EDAPS)*, 2017: IEEE, pp. 1-3.
[51] P. Gaubert, A. Teramoto, and T. Ohmi, "Modelling of the hole mobility in p-channel MOS transistors fabricated on (1 1 0) oriented silicon wafers," *Solid-State Electronics,* doi: 10.1016/j.sse.2009.11.004 vol. 54, no. 4, pp. 420-426, 2010.
[52] F. Larki et al.,. "Simulation of transport in laterally gated junctionless transistors fabricated by local anodization with an atomic force microscope." *physica status solidi (a)* 210.9 (2013): 1914-1919.
[53] F. Larki *et al.*,"Study of carrier velocity of lateral gate p-type silicon nanowire transistor (PSNWT)." *Solid State Science and Technology* 17 (2012).
[54] F. Larki *et al.*, "Pinch-off effect in P-type double gate and single gate junctionless silicon nanowire transistor fabricated by atomic force microscopy nanolithography." *Nano Hybrids* 4 (2013): 33-45.